\documentclass[%
 reprint, amsmath, amssymb, aps,prl
]{revtex4-1}

\usepackage{graphicx}% Include figure files
\usepackage{dcolumn}% Align table columns on decimal point
\usepackage{bm}% bold math
\usepackage{hyperref}% add hypertext capabilities
\usepackage[mathlines]{lineno}% Enable numbering of text and display math

\usepackage{cases}% for dealing with mathematics,
\usepackage{booktabs}
\usepackage{comment}
\usepackage{xcolor}
\usepackage{multirow}
\usepackage{siunitx}
\usepackage[normalem]{ulem}

\newcommand{\Cov}{\mathrm{Cov}}

\newcommand{\ellmin}{\ell_\mathrm{min}}
\newcommand{\ellmax}{\ell_\mathrm{max}}
\begin{document}

\preprint{APS/123-QED}

\title[Sample title]{ 
%A new measurement 
New Extraction of the Cosmic Birefringence from the Planck 2018 Polarization Data}% 

\author{Yuto Minami}
\email{yminami@post.kek.jp}
\affiliation{High Energy Accelerator Research Organization, 1-1 Oho, Tsukuba, Ibaraki 305-0801, Japan}
\author{Eiichiro Komatsu}%
 \email{komatsu@mpa-garching.mpg.de}
\affiliation{ 
Max Planck Institute for Astrophysics, Karl-Schwarzschild-Str. 1, D-85748 Garching, Germany
}%
\affiliation{Kavli Institute for the Physics and Mathematics of the Universe (Kavli IPMU, WPI), Todai Institutes for Advanced Study, The University of Tokyo, Kashiwa 277-8583, Japan}

\date{\today}% It is always \today, today,

\begin{abstract}
We search for evidence of parity-violating physics in the Planck 2018 polarization data, and report on a new measurement of the cosmic birefringence angle, $\beta$.
The previous measurements are limited by the systematic uncertainty in the absolute polarization angles of the Planck detectors.
We mitigate 
this systematic uncertainty completely by simultaneously determining $\beta$ and the angle miscalibration using the observed cross-correlation of the $E$- and $B$-mode polarization of the cosmic microwave background and the Galactic foreground emission.
We show that the systematic errors are effectively mitigated
and achieve a factor-of-$2$ smaller uncertainty than the previous measurement,
finding $\beta=0.35 \pm 0.14\,\deg$ (68\%\,C.L.),
which excludes $\beta = 0$ at $99.2$\%\,C.L. 
This corresponds to the statistical significance of $2.4\sigma$.
\end{abstract}

\maketitle

\section{\label{sec:Introduction}Introduction}
Violation of symmetry in a physical system under parity transformation is sensitive to new physics beyond the standard model (SM) of elementary particles and fields. So far, parity violation has been observed only in the weak interaction~\cite{Lee:1956qn,Wu:1957}.
In the SM of cosmology, called the $\Lambda$ cold dark matter ($\Lambda$CDM) model, the energy budget of the present-day Universe is dominated by unidentified dark matter and dark energy~\cite{Weinberg:2008zzc}.
If dark matter and energy originate from new physics beyond the SM, do either or both of them violate parity?

Polarization of the cosmic microwave background (CMB) is sensitive to parity-violating physics.
Combinations of the Stokes parameters of linear polarization measured in a direction of $\hat{n}$, $Q(\hat{n})\pm iU(\hat{n})$, transform as a spin $\pm 2$ quantity under rotation of $\hat{n}$.
We can use the spin-2 spherical harmonics to decompose these into the so-called $E$- and $B$-mode polarization as
$Q(\hat{n})\pm iU(\hat{n})=-\sum_{\ell m}(E_{\ell m}\pm iB_{\ell m}) {}_{\pm 2}Y_{\ell m}(\hat{n})$
\cite{Seljak:1996gy,Kamionkowski:1996zd}.
Under parity transformation $\hat{n}\to -\hat{n}$, the coefficients transform as $E_{\ell m}\to (-1)^\ell E_{\ell m}$ and $B_{\ell m}\to (-1)^{\ell+1} B_{\ell m}$. When defining angular power spectra as $C_\ell^{AA'}\equiv (2\ell+1)^{-1}\sum_m A_{\ell m}{A'}^*_{\ell m}$ with $A=\{E,B\}$, then $C_\ell^{EE}$ and $C_\ell^{BB}$ are invariant under parity transformation, whereas the cross-power spectrum, $C_\ell^{EB}$, changes the sign. Therefore, nonzero values of $C_\ell^{EB}$ indicate parity violation \cite{Lue:1998mq}.

Pseudoscalar, ``axionlike'' fields, $\phi$, can act as dark matter, energy, or both
(see \cite{Marsh:2015xka,Ferreira:2020fam} for reviews).
A Chern--Simons coupling of a time-dependent $\phi(t)$ to the electromagnetic tensor and its dual, $\frac14g_{\phi\gamma}\phi F_{\mu\nu}\tilde F^{\mu\nu}$, in the Lagrangian density rotates the plane of linear polarization of photons~\cite{Carroll:1989vb,Harari:1992ea,Carroll:1998zi}. This effect, called the ``cosmic birefringence,'' rotates the CMB linear polarization by an angle $\beta=\frac12 g_{\phi\gamma}\int^{t_0}_{t_{\rm LSS}} dt~\dot{\phi}$, and yields a nonzero observed $EB$ spectrum as $C_\ell^{EB, o}=\frac12\sin(4\beta)(C_\ell^{EE}-C_\ell^{BB})$~\cite{Lue:1998mq,Feng:2004mq,Feng:2006dp,Liu:2006uh}, where the subscript ``$o$'' denotes the observed value, the spectra on the right-hand side the intrinsic $EE$ and $BB$ spectra at the last scattering surface (LSS), and $t_0$ and $t_{\rm LSS}$ the times at present and LSS, respectively.

To determine $\beta$, we must know the polarization-sensitive directions of detectors at the focal plane with respect to the sky coordinates. This requires accurate calibration of the polarization angles. Any remaining miscalibration angle, $\alpha$, leads to the same effect as isotropic $\beta$, i.e., $\beta$ and $\alpha$ are degenerate in CMB~\cite{Wu:2008qb,Komatsu:2010fb,Keating:2012ge}.
Recent determinations include 
$\alpha+\beta=-0.36\pm 1.24\,\deg$
from the Wilkinson Microwave Anisotropy Probe (WMAP)~\cite{Hinshaw:2012aka}, $0.31 \pm 0.05\,\deg$ from the Planck mission~\cite{Aghanim:2016fhp},
$-0.61 \pm 0.22\,\deg$ from POLARBEAR~\cite{Adachi:2019mjv},
$0.63 \pm 0.04\,\deg$ from the South Pole Telescope (SPTpol)~\cite{Bianchini:2020osu},
and $0.12 \pm 0.06\,\deg$~\cite{Namikawa:2020ffr} and $0.09 \pm 0.09\,\deg$~\cite{Choi:2020hoge} from
the Atacama Cosmology Telescope (ACT) (also see \cite{Kaufman:2014rpa} for a summary of other experiments).
Here the error bars show the 68\% confidence levels (C.L.) for the statistical uncertainty.
%of $C_\ell^{EB}$ analysis
To isolate $\beta$, an independent estimation of $\alpha$ is required. For  WMAP and Planck the ground calibration yields the systematic uncertainty of $\sigma_{\rm syst}(\alpha)=1.5^\circ$ and $0.28^\circ$, whereas the estimates of systematic uncertainty are not yet available for POLARBEAR, SPTpol, and ACT. 

There is no evidence for nonzero $\beta$ so far.
For the Planck measurement $\sigma_{\rm syst}(\alpha)=0.28^\circ$ is the dominant source of uncertainty for $\beta$. 
How do we make progress in distinguishing between $\beta$ and $\alpha$? In Refs.~\cite{Minami:2019ruj,Minami:2020xfg,MinamiKomatsu:2020} 
we showed that we can simultaneously determine $\alpha$ and $\beta$ if we use the CMB and Galactic foreground emission, as both are rotated by $\alpha$, whereas only the CMB is rotated by $\beta$. Our method thus relies on the different frequency and multipole dependence of the CMB and foreground polarization power spectra. In this Letter, we use this new method to recalibrate the Planck high frequency instrument (HFI) detectors~\cite{PlanckHFI:2018} and measure the cosmic birefringence angle, $\beta$, with a smaller total uncertainty. 

To this end, we assume that there was no intrinsic $EB$ correlation of CMB at the LSS. However, the intrinsic CMB $EB$ can be accounted for if necessary; as such, intrinsic $C_\ell^{EB}$ usually has very different $\ell$ dependence (e.g., \cite{Thorne:2017jft}). For the baseline result we also assume that there is no intrinsic $EB$ correlation of the foreground, but we relax this assumption towards the end of the Letter. 

\section{\label{sec:Map}Maps to cross power spectra}
We use Planck maps from the third public release, referred to as ``PR3''.
We analyze the polarization maps in four polarized Planck HFI channels: $\nu\in \{100, 143, 217, 353\}\,\si{\GHz}$. We also use the temperature maps when we correct the temperature-to-polarization ($I \to P$) leakage effect due to beams.
We cross-correlate four frequency maps from different half-mission (HM) maps, HM1 and HM2,
to reduce the correlated systematics and bias from the auto correlation noise.

To reject spurious signals, we apply three types of masks.
(1) Bad pixels: we remove the pixels that were not observed by any detectors.
(2) Bright CO emission: the Planck team used the bandpass templates to correct for CO emission, which were generated at $N_\mathrm{side}=128$ in the HEALPix format \cite{Gorski:2004by}. The difference between this and the native resolution of the HM maps ($N_\mathrm{side}=2048$) causes a bias, which is significant in bright CO emission regions. 
To reduce the bias, we follow Planck team's suggestion and mask the bright CO regions where the bias level is larger than $1\%$ of the noise level~\cite{PlanckHFI:2018}.
We have applied this mask to all channels except for $143\,$GHz channel, to which no CO bandpass template was applied.
(3) Bright point sources: we use the point-source mask provided by the Planck team, which removes sources with polarization detection significance levels of $\geq99.97\%$.

We apply the combined masks to the HM maps. 
We then estimate observed power spectra, $C_\ell^{XY, \mathrm{o}}$,
with $XY \in \{TT, EE, BB, TE,ET, EB, BE\}$
from  16 combinations of the masked HM maps
using the \textsc{NaMaster} package~\cite{Alonso:2018jzx}.
When estimating $C_\ell^{XY, o}$ we apodize the combined masks with $0.5\deg$ using the ``Smooth'' method of \textsc{NaMaster}. The fractions of sky used for the analysis are calculated as $f_\mathrm{sky}=\sum_{i=1}^{N_{\rm pix}} w_i^2/N_{\rm pix}$, where $w_i$ is the value of (non-integer) smoothed mask and $N_{\rm pix}=12 N_\mathrm{side}^2$ is the number of pixels of the HM maps. We find $(f_\mathrm{sky}^{\nu,\mathrm{HM1}},~f_\mathrm{sky}^{\nu,\mathrm{HM2}})=\{(0.97, 0.95),(0.94, 0.90), (0.82, 0.77), (0.92, 0.89 )\}$ for $\nu \in \{100, 143, 217, 353\} \, \si{\GHz}$, respectively.

To remove the  $I \to P$ leakage, we use the beam window matrix, $ W_\ell^{XY,X'Y'}$, produced by the ``QuickPol'' method~\cite{Hivon:2016qyw}.
The matrix describes how the observed $XY$ power spectra
are related to the input ones with $X' Y' \in \{TT,EE,BB,TE\}$.
Since our power spectra include both the CMB and foregrounds, we do not have a prior knowledge of the input power spectra.
Therefore, we approximately use the \textit{observed} power spectra divided by the diagonal elements of the beam window matrix as the input. In summary,
the observed power spectra after the leakage subtraction are given by 
\begin{align}
&C_\ell^{XY,\mathrm{o}} =\\ \nonumber
\ &\hat{C}_\ell^{XY,\mathrm{o}} 
- W_{\ell}^{\mathrm{pix},XY}\sum_{X'Y'\neq XY}  
\frac{ W_\ell^{XY,X'Y'} \hat{C}_\ell^{X'Y',\mathrm{o}} }{W_\ell^{\mathrm{pix},X'Y'} W_\ell^{X'Y',X'Y'}}
,
\end{align}
where $\hat{C}_\ell^{XY}$ is a power spectrum before the leakage subtraction,
and $W_\ell^{\mathrm{pix},XY}$ is a pixel window function for the $XY$ power spectrum.
Because QuickPol assumes that the signal is statistically isotropic on the sky,
the leakage from $ET$ is equal to that from $TE$; thus, 
we use the mean of $TE$ and $ET$ as an input for $X' Y' = TE$. 

\section{\label{sec:Methodology}Estimation of \texorpdfstring{$\alpha$ and $\beta$}{alpha and beta}}
We estimate one global cosmic birefringence angle, $\beta$, and independent miscalibration angles, $\alpha_{\nu}$, at four frequencies. 
When the intrinsic $EB$ power spectra of the CMB at LSS and the Galactic foregrounds vanish,
we can relate the observed power spectra and the best-fitting $\Lambda$CDM CMB power spectra \footnote{We use the CMB power spectra calculated by CAMB~\cite{Lewis:2000}
using the Planck 2018 cosmological parameters for ``TT,TE,EE$+$lowE$+$lensing''~\cite{Aghanim:2018eyx}:
$\Omega_bh^2=0.022\,37$, $\Omega_ch^2=0.1200$,
$h=0.6736$, $\tau=0.0544$, $A_s=2.100\times 10^{-9}$, and $n_s=0.9649$.} at each $\ell$ as~\cite{MinamiKomatsu:2020}
\begin{align}
\mathbf{A}\vec C_\ell^{\mathrm{o}} - \mathbf{B
}\vec C_\ell^\mathrm{CMB,th} = \mathbf{0},
\end{align}
where $\vec{C_\ell^{\mathrm{o}}}$ is an array of the observed power spectra,
$\begin{pmatrix}
C_\ell^{E_i E_j,\mathrm{o}}&
C_\ell^{B_i B_j,\mathrm{o}} &
C_\ell^{E_i B_j,\mathrm{o}}
\end{pmatrix}^T,$ with $i,j$ in $32$ combinations,
$\vec C_\ell^\mathrm{CMB,th}$ is an array of the best-fitting $\Lambda$CDM CMB power spectra,
$\left(
C_\ell^{E_i E_j,\mathrm{CMB,th}} W_\ell^{E_i E_j,E_i E_j}W_\ell^{\mathrm{pix},E_i E_j}\right.$
$\left. 
C_\ell^{B_i B_j,\mathrm{CMB,th}} W_\ell^{B_i B_j, B_i B_j}W_\ell^{\mathrm{pix},B_i B_j}
\right)^T,$
with the corresponding beam window matrix,
$\mathbf{A}$ is a block diagonal matrix of $\begin{pmatrix}
- \vec{R}^T(\alpha_i,\alpha_j)\mathbf{R}^{-1} (\alpha_i,\alpha_j)& 1
\end{pmatrix}$,
and $ \mathbf{B}$ is a block diagonal matrix of $ \left( \vec{R}^T ( \alpha_i+\beta, \alpha_j+\beta)\right.$  $\left.- \vec{R}^T(\alpha_i,\alpha_j) \mathbf{R}^{-1}(\alpha_i,\alpha_j) \mathbf{R}(\alpha_i+\beta, \alpha_j+\beta) \right)$.
Here, $\mathbf{R}$ and $ \vec{R}$ are the rotation matrix and vector defined in  Eq.~(8) and (9) of Ref.~\cite{MinamiKomatsu:2020}, respectively.
We have 32 independent equations from 16 combinations of maps,
as we have two different equations for $C_\ell^{E_iB_j,\mathrm{o}}$ and $C_\ell^{E_jB_i,\mathrm{o}}$.

In practice, we estimate $\alpha_\nu$ and $\beta$ by maximizing the log-likelihood function \cite{MinamiKomatsu:2020}:
\begin{align}\label{eq:LikelihoodCross}
&\ln\mathcal{L} =
-\frac{1}{2}
\sum_{\ell=\ellmin}^{\ellmax}
\vec{v}^T_\ell 
\mathbf{C}^{-1}_\ell
\vec{v}_\ell,
\end{align}
where $\vec{v}_\ell\equiv \mathbf{A}\vec C_\ell^{\mathrm{o}} - \mathbf{B}\vec C_\ell^\mathrm{CMB,th}$ and $\mathbf{C}_\ell \equiv \mathbf{A}\Cov(\vec C_\ell^{\mathrm{o}},\vec C_\ell^{\mathrm{o}}{}^T)\mathbf{A}^T$.
We use a publicly available Markov chain Monte Carlo sampler \textsc{emcee}~\cite{ForemanMackey:2012ig} to obtain posterior distributions of $\alpha_\nu$ and $\beta$ with this likelihood and flat priors on $\alpha_\nu$ and $\beta$.
As we estimate the covariance matrix from the observed power spectra, we use binned power spectra with $\Delta\ell=20$ to reduce the statistical fluctuation in the covariance matrix. 
We follow the definition of $\Cov(\vec C_\ell^{\mathrm{o}},\vec C_\ell^{\mathrm{o}}{}^T)$ given in Eqs.~(12)-(15) of Ref.~\cite{MinamiKomatsu:2020}, but with a slight modification to account for the effect of mask. Specifically, we divide the covariance matrix by $f_\mathrm{sky}^\mathrm{eff} = \sqrt[4]{f_\mathrm{sky}^{i}f_\mathrm{sky}^{j}f_\mathrm{sky}^{p}f_\mathrm{sky}^{q}}$ with $f_\mathrm{sky}^{i}$ being $f_\mathrm{sky}$ for the $i$th map.

Our covariance matrix formula is valid for approximately Gaussian random fields;
however,  non-Gaussian effects from, e.g., the foreground, may become non-negligible at low multipoles.
To find a suitable minimum multipole, $\ellmin$, we vary $\ellmin$ from $2$ to $200$ and estimate $\alpha_\nu$ and $\beta$.
We obtain stable results for $\ellmin\approx50$. Specifically, we find $\beta=0.71\pm 0.14$ and $ 0.48\pm 0.14$ deg for $\ellmin = 25$ and $41$, respectively, but then find a stable value of $\beta=0.35\,\deg$ to within the uncertainty for $\ellmin\gtrsim 50$;
thus, we use $\ellmin=51$, which coincides with the value adopted by the Planck team~\cite{Aghanim:2016fhp}.

As for the maximum multipole, $\ellmax$, we use the same $\ellmax=1500$ as in the Planck analysis~\cite{Aghanim:2016fhp}.

\section{\label{sec:Validation}Validation with the full focal plane simulation}
To validate our pipeline, 
we first use the maps from Planck's end-to-end full focal plane 10 (FFP10) simulation \cite{PlanckHFI:2018}.
Since the FFP10 simulation does not have foreground maps convolved with realistic beam effects such as the $I \to P$ leakage, we only consider CMB and noise realizations of the HM maps.

As the maps do not include the foreground, we can only estimate the combination $\alpha_\nu+\beta$. 
Thus, we estimate (i) $\alpha_\nu$ by setting $\beta=0\,\deg$ and (ii) $\beta$ by setting $\alpha_\nu =0\,\deg$ for 10 realizations.
We expect to recover (i) $\alpha_\nu=0$ and (ii) $\beta=0$, as the FFP10 simulation does not include angle miscalibration or the cosmic birefringence. 
The means and standard deviations of the recovered angles are (i) $\alpha_\nu = \{-0.008\pm0.047,0.013\pm0.033, 0.017\pm0.065, 0.14\pm0.41\}\,\deg$ for $\nu\in\{100,143, 217, 353\}\,\si{GHz}$
 and (ii) $\beta = 0.010 \pm 0.030\,\deg$.
We thus find no evidence for a spurious $\alpha_\nu$ or $\beta$ from the instrumental effects, to the extent that is implemented in the FFP10 simulation.

\section{\label{sec:Results}Results}
First, we assume that the polarization directions of the Planck detectors are perfectly calibrated, i.e., $\alpha_\nu=0$, and estimate $\beta$. This case is similar to the Planck analysis \cite{Aghanim:2016fhp}, except that they measured $\beta$ from foreground-cleaned maps. 
We find $\beta(\alpha_\nu=0)=0.289\pm0.048\,\deg$,
which is consistent with the Planck team's result,
$0.29 \pm 0.05\,\mathrm{(stat.)} \pm 0.28\,\mathrm{(syst.)}$ from $C_\ell^{EB}$, within the statistical uncertainty.
When $C_\ell^{TB}$ is added they find $0.31\,\deg$.
The second error bar of the Planck measurement is the systematic uncertainty in $\alpha$ from the ground calibration.
Our goal is to estimate $\alpha_\nu$ simultaneously to eliminate this uncertainty. 
Nevertheless, it is reassuring that we obtain consistent results under a similar setup. 

\begin{figure}[tp]
\centering
\includegraphics[width=0.95\linewidth]{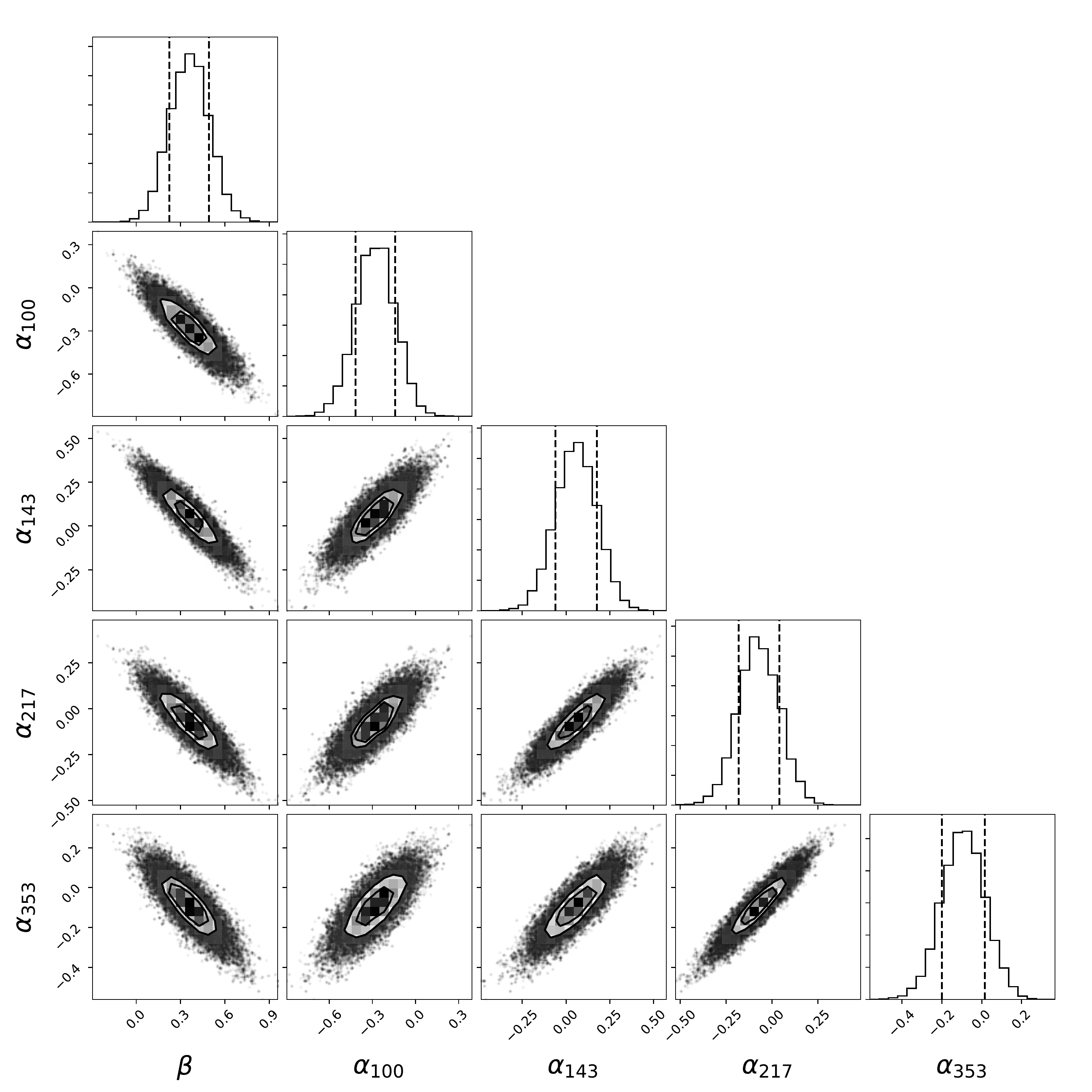}
\caption{\label{fig:PostDist} Posterior distributions of
$\beta$ (the first column) against the miscalibration angles $\alpha_\nu$. 
The solid contour lines in the 2D histograms show 1$\sigma$ ($39.3\%$) and 2$\sigma$ ($86.5\%$) of each area.
The dashed lines in the 1D histograms show $1\sigma$ (from $16\%$ to $84\%$) quantiles of each area.}
\end{figure}

Next, we estimate $\beta$ and $\alpha_\nu$ simultaneously. We report our baseline results in Table~\ref{tab:result}, and the posterior distributions of the angles in Fig.~\ref{fig:PostDist}.
It shows that $\alpha_\nu$ and $\beta$ are anticorrelated, since the CMB determines $\alpha_\nu+\beta$ and the degeneracy is broken by the foregrounds \cite{Minami:2019ruj}.
We find that the miscalibration angles are consistent with zero to within 1$\sigma$ at 143, 217, and 353~GHz, and is a 2$\sigma$ level at 100~GHz. All the values are within the systematic uncertainty of the ground calibration, $\sigma_{\rm syst}(\alpha)=0.28$\,deg. Our baseline result is $\beta=0.35\pm 0.14$\,deg, which excludes the null hypothesis by $99.2$\%~C.L. The uncertainty no longer contains the ground calibration uncertainty, as we
simultaneously determine $\alpha_\nu$ and $\beta$.
Our measurement is consistent with the Planck team's result quoted above, with a factor-of-$2$ smaller total uncertainty. 

\begin{table}[tp]%The best place to locate the table environment is directly after its first reference in text
	\caption{\label{tab:result}% 
		Cosmic birefringence and miscalibration angles from the Planck 2018 polarization data with $1\sigma~(68\%)$ uncertainties
	}
	\begin{ruledtabular}
		\begin{tabular}{cc}
			Angles &\textrm{Results ($\deg$)}\\
			\colrule
			$\beta$  & $0.35\pm0.14$ \\
			$\alpha_{100}$ &  $-0.28 \pm 0.13$\\
			$\alpha_{143}$ &   $ 0.07 \pm 0.12$\\
			$\alpha_{217}$ &  $-0.07 \pm 0.11$ \\
			$\alpha_{353}$ &   $-0.09 \pm 0.11$\\
		\end{tabular}
	\end{ruledtabular}
\end{table}

\begin{figure*}[tp]
	\centering
	\includegraphics[width=\linewidth]{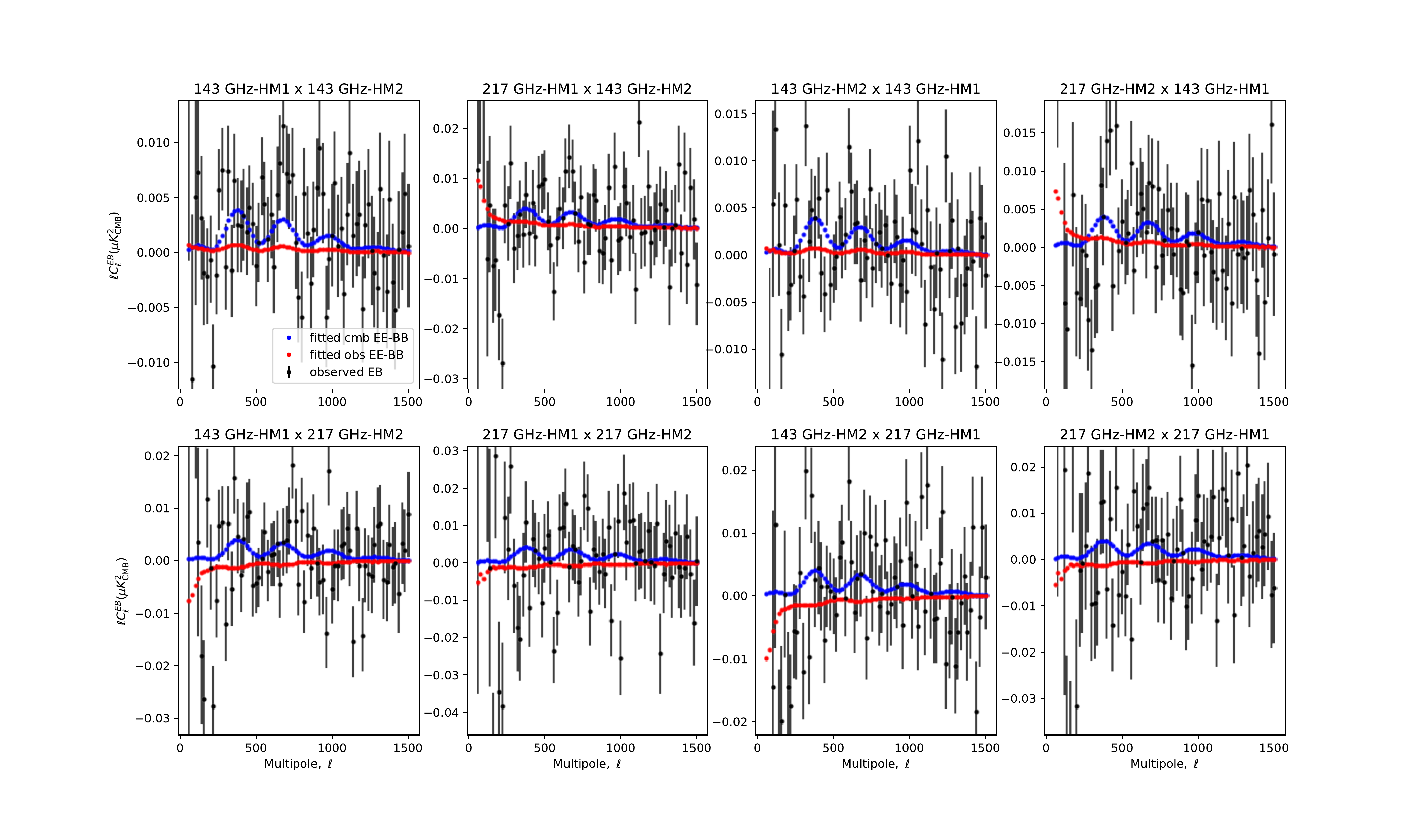}
	\caption{\label{fig:EBplot} Fitted $EB$ cross spectra from $143$ and $217$\,\si{\GHz} maps. We show the measured $EB$ data with error bars (black),
	$C_\ell^{EE} - C_\ell^{BB}$ terms of observed $-\mathbf{A}\vec C_\ell^{\mathrm{o}}$ (red),
	and the CMB $\mathbf{B}\vec C_\ell^\mathrm{CMB,th}$ (blue).
    The data points should be compared with the sum of $C_\ell^{EE} - C_\ell^{BB}$ terms.
	}
\end{figure*}

We show the fitted $EB$ power spectra of $143$ and $217$\,\si{\GHz}, which
%have small error bars among all the frequencies
have the smallest error bars,
in Fig.~\ref{fig:EBplot}.
The measured data points with error bars should be compared with the sum of $C_\ell^{EE} - C_\ell^{BB}$ terms of
$-\mathbf{A}\vec C_\ell^{\mathrm{o}}$ (red) and $\mathbf{B
}\vec C_\ell^\mathrm{CMB,th}$ (blue).
\begin{comment}
$C_\ell^{EE,\mathrm{o}}-C_\ell^{BB,\mathrm{o}}$ rotated by the best-fitting $\alpha_\nu$ (red) and $C_\ell^{EE,\mathrm{CMB,th}}-C_\ell^{BB,\mathrm{CMB,th}}$ rotated by $\alpha_\nu+\beta$ (blue).
\end{comment}
To guide eyes, we note that  the $217\,\si{\GHz}$-HM1$\times143\,\si{\GHz}$-HM2 panel shows the $EB$ power spectrum with a hint of the acoustic oscillation matched by the CMB $E$-mode power spectrum. Similar trends are seen in some of the other panels, explaining a 2.4$\sigma$ hint for a nonzero value of $\beta$.

While it is perfectly consistent with the quoted systematic uncertainty of the ground calibration, one may wonder if $\alpha_{100}=-0.28\pm 0.13$\,deg is the cause for a nonzero value of $\beta$.
One potential source of worry is the $EB$ correlation of synchrotron radiation which may become important at lower $\nu$.
The intrinsic $EB$ correlation of synchrotron, if any, may create the bias.
To test this, we exclude the $100\,$GHz channel and repeat the analysis. We find $\beta= 0.40\pm 0.15\,\deg$ and
$\alpha_\nu = \{0.05 \pm 0.12, -0.13 \pm 0.12,  -0.10\pm0.11\}\,\deg$ for $\nu\in\{143, 217, 353\}\,\si{GHz}$, which agree with the baseline. 

We test the effect of the
$I\to P$ leakage  by estimating $\alpha_\nu$ and $\beta$ without the leakage subtraction.
We find $\beta= 0.35\pm 0.14\,\deg$ and
$\alpha_\nu = \{-0.25 \pm 0.14, 0.07 \pm 0.12, -0.05 \pm 0.11,  -0.07\pm0.11\}\,\deg$ for $\nu\in\{100,143, 217, 353\}\,\si{GHz}$, which agree with the baseline; thus, the results are robust against the leakage. 

\section{\label{sec:FGEB}\texorpdfstring{$EB$}{EB} correlation from the Galactic foreground}

So far, we have assumed that the intrinsic $EB$ power spectrum of the foreground emission vanishes. In this section we relax this assumption. In the previous section we have shown that dropping the 100~GHz channel does not affect the result for $\beta$ \footnote{We further check the non-importance of the synchrotron foreground by computing $EB$ due to synchrotron for $\SI{143}{\GHz}$-HM1$\times\SI{217}{\GHz}$-HM2, where variance of the observed $EB$ is small. We estimate a Gaussian variance of synchrotron $EB$ using a synchrotron model implemented in the public code ``PySM''~\cite{Thorne:2016ifb}.
The ratio of the synchrotron variance to the observed variance is $O(10^{-9})$.
Thus, even if synchrotron has a significant $EB$ correlation at the level of $5\sigma$, the effect is negligible.}.
Therefore, we focus on the dust emission, which is the dominant foreground in the Planck HFI channels. 

As discussed in Refs.~\cite{Minami:2019ruj,Minami:2020xfg},
we can parameterize the dust $EB$ power spectrum by a frequency-dependent rotation angle, $\gamma(\nu)$, as 
$ C_\ell^{EB,\mathrm{dust}}  = \frac{\sin\left[4\gamma(\nu) \right]}{2} \left( C_\ell^{EE,\mathrm{dust}}  -  C_\ell^{BB,\mathrm{dust}} \right)$.
The sign of the $EB$ correlation is the same as $\gamma$ because $C_\ell^{EE,\mathrm{dust}}>C_\ell^{BB,\mathrm{dust}}$~\cite{planckdust:2018}.
In the worst case scenario $\gamma$ is independent of frequency, which would make it indistinguishable from $\beta$. Then, our result can be reinterpreted as the combination of angles $\beta -\gamma= 0.35 \pm 0.14\,\deg$.
Because both the $TE$ and $TB$ cross power spectra of thermal dust emission are positive~\cite{planckdust:2018},
a positive $EB$, hence $\gamma>0$, is expected;
thus,  our baseline result assuming $\gamma=0$ gives a lower bound for $\beta$.

What if $\gamma<0$? If all of the signal we see in $\beta$ is due to the dust emission, it implies $\gamma = -0.35 \pm 0.14\,\deg$.
In this case,
assuming $\xi = C_\ell^{BB,\mathrm{dust}}/C_\ell^{EE,\mathrm{dust}}\simeq0.5$~\cite{planckdust:2018,Abitbol:2015epq},
we find a correlation coefficient of $f_c = C_\ell^{EB,\mathrm{dust}}/\sqrt{ C_\ell^{EE,\mathrm{dust}} C_\ell^{BB,\mathrm{dust}}} \simeq (-8.6 \pm 3.5) \times 10^{-3}$, whose absolute value corresponds to the lowest value of $f_c$ discussed in Ref.~\cite{Abitbol:2015epq}.

\section{\label{sec:Conclusion}Summary and discussion}
In this Letter, we have applied the new method of simultaneously determining the cosmic birefringence angle $\beta$ and miscalibration angles of detectors $\alpha_\nu$ to the Planck 2018 data. The method was developed originally in Ref.~\cite{Minami:2019ruj} for autofrequency power spectra measured over the full sky, and has been extended to include a partial sky coverage \cite{Minami:2020xfg} and cross-frequency spectra \cite{MinamiKomatsu:2020}. The idea is simple: while $\alpha_\nu$ rotates linear polarization of both the CMB and Galactic foreground emission, $\beta$ rotates only the CMB. We find that all of $\alpha_\nu$ in the polarized Planck HFI channels are consistent with zero to within the quoted systematic uncertainty of the ground calibration of the Planck bolometers~\cite{Aghanim:2016fhp}.

We measure $\beta=0.35\pm0.14\,\deg$ (68\% C.L.), which excludes zero by 99.2\% C.L.
This corresponds to the statistical significance of $2.4\sigma$.
This value is consistent with the Planck team's result assuming $\alpha_\nu=0$, but with a factor-of-$2$ smaller total uncertainty because our result is no longer subject to the ground calibration uncertainty.

We can constrain various models of new physics which produce a spatially uniform $\beta$.
Let us consider a Lagrangian density including a Chern--Simons coupling between axionlike particles and photons (see, e.g., \cite{Turner:1987bw}):
\begin{align}
\mathcal{L} \supset 
\frac{1}{4} g_{\phi\gamma}\phi F_{\mu\nu}\tilde F^{\mu\nu}\,,
\end{align}
where $g_{\phi\gamma}$ is a coupling constant, $\phi$ is an axionlike pseudoscalar field,
and $F_{\mu\nu}$ and $\tilde{F}_{\mu\nu}$ are the electromagnetic tensor and its dual. 
The difference of the value of $\phi$ between the LSS and the location of the observer (``obs'') rotates the plane of linear polarization of CMB photons by
$\beta = \frac12g_{\phi\gamma}( \bar{\phi}_\mathrm{obs} - \bar{\phi}_\mathrm{LSS} + \delta\phi_\mathrm{obs})$
\cite{Carroll:1989vb,Harari:1992ea,Carroll:1998zi,Lue:1998mq,Feng:2004mq,Feng:2006dp,Liu:2006uh,FujitaMinami:2020},
where $\bar{\phi}$ and $\delta\phi$ denote the mean and fluctuation of the field value, respectively.
Then our measurement gives 
\begin{equation}
g_{\phi\gamma}( \bar{\phi}_\mathrm{obs} - \bar{\phi}_\mathrm{LSS} + \delta\phi_\mathrm{obs}) = \SI[separate-uncertainty = true]{1.2(5)e-2}{\radian}\,.
\end{equation}
We can use this to constrain models (see, e.g., \cite{FujitaMinami:2020}).

If our measurement of $\beta$ is confirmed with higher statistical significance in future, it would have a profound implication for  fundamental physics. To further test and improve our measurement, one can apply our method to both the ongoing \cite{Ade:2014afa,Ade:2014gua,Benson:2014qhw,Xu:2019rne,Choi:2020hoge}
and future \cite{Ben:2018,Hui:2018cvg,Ade:2018sbj,abazajian2019cmbs4,Hazumi2019}
CMB polarization experiments.

\begin{acknowledgments}
We acknowledge the use of the public Planck data released via the Planck Legacy Archive. We thank E. Hivon for his help with the QuickPol beam window matrices, and H.K. Eriksen, M. L{\`o}pez-Caniego, A. Banday, and A, Gruppuso for their help with the $EB$ spectra from the Planck data. We also thank Y. Chinone, K. Ichiki, N. Katayama, T. Matsumura, H. Ochi, and S. Takakura for useful discussions.
Y.~M. thanks T. Fujita,  K. Murai, and H. Nakatsuka for discussion on the cosmic birefringence by axionlike particles.
This work was supported in part by the Japan Society for the Promotion of Science (JSPS) KAKENHI, Grants No.~JP20K1449 and No.~JP15H05896, and the Excellence Cluster ORIGINS which is funded by the Deutsche Forschungsgemeinschaft (DFG, German Research Foundation) under Germany’s Excellence Strategy: Grant No.~EXC-2094 - 390783311.
The Kavli IPMU is supported by World Premier International Research Center Initiative (WPI), MEXT, Japan.
\end{acknowledgments}

\bibliographystyle{apsrev4-1}
\bibliography{references}% Produces the bibliography via BibTeX.

\end{document}